\documentclass[aps,twocolumn]{revtex4}

\usepackage{enumerate}

\usepackage{amsmath}

\begin{document}

\title{Comment on "Connection between the Burgers equation with an elastic  forcing term and a stochastic process"}
\author{Piotr Garbaczewski\thanks{electronic address: p.garbaczewski@if.uz.zgora.pl}}
\affiliation{Institute of Physics,  University  of Zielona
G\'{o}ra, 65-516 Zielona G\'{o}ra, Poland}

\begin{abstract}
In the  above mentioned paper by E.  Moreau and O. Vall\'{e}e
[Phys. Rev.  {\bf E 73}, 016112, (2006)], the one-dimensional
Burgers equation with an elastic (attractive) forcing term  has
been claimed to  be connected with the Ornstein-Uhlenbeck process.
We point out that this connection  is  valid only in case of the
repulsive forcing.
\end{abstract}
\maketitle

PACS numbers: 02.50.Ey, 05.90.+m, 05.45.-a
 \vskip0.3cm

Let us consider the  Langevin equation for the one-dimensional stochastic
process in the external conservative  force field $F(x) = - dV(x)/dx$:
\begin{equation}
{\frac{dx}{dt}} = F(x) + \sqrt{2\nu } b(t)
\end{equation}
 where $b(t)$ stands for the normalized white noise: $\langle b(t)\rangle =0$,
 $\langle b(t')b(t)\rangle= \delta (t-t')$. The corresponding Fokker-Planck equation
 for the probability density  $\rho (x,t)$ reads:
 \begin{equation}
 \partial _t\rho = \nu \partial _{xx}\rho  -\partial _x( F\rho)
 \end{equation}
 and  by  means of a standard substitution  $\rho (x,t) = \Psi (x,t) \exp[- V(x)/2\nu]$, \cite{risken}, can be
 transformed into the  generalized diffusion equation for an auxiliary function $\Psi (x,t)$:
 \begin{equation}
 \partial _t \Psi = \nu \partial _{xx}\Psi - {\cal{V}} (x) \Psi
 \end{equation}
 where
 \begin{equation}
 {\cal{V}} (x) = {\frac{1}2}\left( {\frac{F^2}{2\nu }} + \partial _x F\right) \, .
 \end{equation}

As discussed in detail in Refs. \cite{gar,gar1,muratore}, given the so-called  forward  drift  $b(x,t)$ of the
Markovian diffusion process, in the above   identified with $b(x,t) \doteq F(x)$, one readily  infers the
so-called backward drift of this process
\begin{equation}
b_{\star }(x,t) \doteq  b(x,t) -  2\nu \partial _x (\ln \rho )(x)
\end{equation}
which is known to solve the forced Burgers equation:
\begin{equation}
\partial _tb_{\star } + b_{\star } \partial _x b_{\star } - \nu \partial _{xx} b_{\star } = {\cal{F}} \, .
\end{equation}
where $ {\cal{F}}= +  2\nu
\partial _x {\cal{V}} $.

For the Ornstein-Uhlenbeck process  $b(x) = F(x) = - \kappa x$ and accordingly
\begin{equation}
  {\cal{V}} (x) = {\frac{\kappa ^2x^2}{4\nu }} - {\frac{\kappa }2} \, .
\end{equation}
Substituting  the inferred    ${\cal{V}}(x)$  into Eq. (3), we get
Eq. (32) of Ref. \cite{moreau}. We observe that the velocity field
$u(x,t)$,  defined by  Eq. (35) of Ref. \cite{moreau}, does
coincide with our $b_{\star }(x,t)$, provided we set $b(x,t)= -
\kappa x$ in Eq. (5).

 The related ${\cal{V}} (x)$
gives rise to
\begin{equation}
{\cal{F}}(x) =  2\nu  \partial _x{\cal{V}} (x)= + \kappa ^2 x  \,
,
\end{equation}
 while an  original
 elastic forcing  problem addressed by \cite{moreau}, Eq. (2) therein,  has the form:
\begin{equation}
\partial_tu + u\partial _xu - \nu \partial _{xx}u = - \kappa ^2 x
\end{equation}
and clearly  differs from  Eq. (6), with  the necessarily arising $+ \kappa ^2 x $ on its right-hand-side,
 by an innocent-looking but crucial in the present context sign of the forcing term.

Let us add that this particular sign issue has received due  attention in  Ref. \cite{gar2}.
A specific  class of  diffusion-type processes has been  considered that would  account for
standard Newtonian accelerations (of the  form  $-\partial _x {\cal{}}W$ with $W(x)$ bounded
from below) on the right-hand-side  of Eq. (6). It is  in principle  possible at  the price of  introducing
an additional pressure-type  forcing term.

{\bf  Acknowledgement:} This note  has been supported by the Polish Ministry of Scientific Research and
Information Technology under the  (solicited) grant No PBZ-MIN-008/P03/2003.

\end{document}